\documentclass[12pt]{article}

\usepackage{axodraw}
\usepackage{epsf}
\usepackage{rotating}

\def\arraystretch{1.2}

\def\text{\textstyle}

\newcommand{\lsim}{\raisebox{-0.13cm}{~\shortstack{$<$ \\[-0.07cm] $\sim$}}~}

\newcommand{\TeV}{\unskip\,\mathrm{TeV}}
\newcommand{\GeV}{\unskip\,\mathrm{GeV}}

\newcommand{\fb}{\unskip\,\mathrm{fb}}

\def\draftdate{\relax}
\def\mda{\relax}
\def\mua{\relax}
\def\mla{\relax}
\def\draft{
\def\thtystars{******************************}
\def\sixtystars{\thtystars\thtystars}
\typeout{}
\typeout{\sixtystars**}
\typeout{* Draft mode!
         For final version remove \protect\draft\space in source file *}
\typeout{\sixtystars**}
\typeout{}
\def\draftdate{\today}
\def\mua{\marginpar[\boldmath\hfil$\uparrow$]%
                   {\boldmath$\uparrow$\hfil}%
                    \typeout{marginpar: $\uparrow$}\ignorespaces}
\def\mda{\marginpar[\boldmath\hfil$\downarrow$]%
                   {\boldmath$\downarrow$\hfil}%
                    \typeout{marginpar: $\downarrow$}\ignorespaces}
\def\mla{\marginpar[\boldmath\hfil$\rightarrow$]%
                   {\boldmath$\leftarrow $\hfil}%
                    \typeout{marginpar: $\leftrightarrow$}\ignorespaces}
\overfullrule 5pt
\oddsidemargin -15mm
\marginparwidth 29mm
}

\catcode`\@=11
\ifx\@Hxfloat\@Hundef\else\expandafter\endinput\fi
\let\@Hxfloat\@xfloat
\def\@xfloat#1[{\@ifnextchar{H}{\@HHfloat{#1}[}{\@Hxfloat{#1}[}}
\def\@HHfloat#1[H]{%
\expandafter\let\csname end#1\endcsname\end@Hfloat
\vskip\intextsep\vbox\bgroup\def\@captype{#1}\parindent\z@
\ignorespaces}
\def\end@Hfloat{\egroup\vskip \intextsep}
\def\section{\@startsection{section}{1}{\z@}{3.5ex plus 1ex minus .2ex}
{2.3ex plus .2ex}{\large\bf}}
\def\thesection{\arabic{section}.}

\def\appendix{\setcounter{section}{0}
 \def\thesection{APPENDIX \Alph{section}:}
 \def\theequation{\Alph{section}.\arabic{equation}}}
\def\@citex[#1]#2{\if@filesw\immediate\write\@auxout{\string\citation{#2}}\fi
  \def\@citea{}\@cite{\@for\@citeb:=#2\do
    {\@citea\def\@citea{,\penalty\@m}\@ifundefined
       {b@\@citeb}{{\bf ?}\@warning
       {Citation `\@citeb' on page \thepage \space undefined}}%
\hbox{\csname b@\@citeb\endcsname}}}{#1}}

\def\citer{\@ifnextchar [{\@tempswatrue\@citexr}{\@tempswafalse\@citexr[]}}
\def\@citexr[#1]#2{\if@filesw\immediate\write\@auxout{\string\citation{#2}}\fi
  \def\@citea{}\@cite{\@for\@citeb:=#2\do
    {\@citea\def\@citea{--\penalty\@m}\@ifundefined
       {b@\@citeb}{{\bf ?}\@warning
       {Citation `\@citeb' on page \thepage \space undefined}}%
\hbox{\csname b@\@citeb\endcsname}}}{#1}}
\begin{document}

\renewcommand{\thefootnote}{\fnsymbol{footnote}}
\setcounter{page}{0}

\begin{titlepage}

\begin{flushright}
DESY 99-177 \\
BI-TP 99/27 \\
Edinburgh 2000/03 \\
hep-ph/0002035
\end{flushright}

\vspace*{1.0cm}

\begin{center}
{\large \bf HIGGS RADIATION OFF QUARKS} \\[0.5cm]
{\large \bf IN SUPERSYMMETRIC THEORIES AT \boldmath{$e^+e^-$} COLLIDERS}\\
\end{center}

\vskip 1.cm
\begin{center}
{\sc S. Dittmaier$^1$, M. Kr\"amer$^2$\footnote{Supported by the EU FF
Programme under contract FMRX-CT98-0194 (DG 12 - MIHT).}, Y. Liao$^{3,4}$,}
\\[0.3cm]
{\sc M. Spira$^5$\footnote{Heisenberg-Fellow} and P.M. Zerwas$^4$}

\vskip 0.8cm

\begin{small} 
$^1$ Theoretische Physik, Universit\"at Bielefeld,
D-33615 Bielefeld, Germany
\vskip 0.2cm
$^2$ Dept.\ of Physics and Astronomy, Univ.\ of Edinburgh,
Edinburgh EH9 3JZ, Scotland
\vskip 0.2cm
$^3$ Department of Modern Applied Physics, Tsinghua University, 
Beijing 100084, PR China
\vskip 0.2cm
$^4$  DESY, Deutsches Elektronen-Synchrotron, D-22603 Hamburg, Germany 
\vskip 0.2cm
$^5$ II.\ Institut f\"ur Theoretische Physik$^\ast$,
Universit\"at Hamburg, D-22761 Hamburg, Germany
\end{small}
\end{center}

\vskip 2cm

\begin{abstract}
\noindent
Yukawa couplings between Higgs bosons and quarks in supersymmetric
theories can be measured in the processes $e^+e^-\to Q\bar{Q} +
\mbox{Higgs}$.  We have determined the cross sections of these
processes in the minimal supersymmetric model including the complete
set of next-to-leading order QCD corrections for all channels.
\end{abstract}

\end{titlepage}

\renewcommand{\thefootnote}{\arabic{footnote}}

\setcounter{footnote}{0}

\clearpage

\noindent
{\bf 1.} In the Standard Model (SM) and its supersymmetric extensions, the
masses of electroweak gauge bosons, leptons, and quarks are generated
by interactions with Higgs fields \cite{hi64}. The Yukawa couplings
between Higgs particles and fermions therefore grow with the masses
$M_f$ of the fermions. The couplings obey a universal scaling law
$g_{ffH}=M_f/v$ in the SM, with $v\approx 246\GeV$ being the
ground-state value of the Higgs field. In supersymmetric models, which
involve at least two Higgs doublets, the size of the Yukawa couplings
is also set by the fermion masses, yet the relationship is more
complex owing to the mixing among the Higgs fields. The Yukawa couplings
\cite{gu86} of the two CP-even light/heavy Higgs bosons $h/H$
and of the CP-odd Higgs boson $A$ in the minimal supersymmetric
extension of the Standard Model (MSSM) \cite{mssma,mssmb}, expressed in
units of the SM couplings, are collected in Table~\ref{tab:coeffs}%
\footnote{For large $\tan\beta$, SUSY loop corrections to the 
$b\bar b$--Higgs vertices become very important \cite{car98}.  In
the present QCD analysis, these corrections can be absorbed
into effective Higgs Yukawa couplings, since emission and reabsorption
of virtual heavy SUSY particles is confined to small space-time
regions compared with QCD subprocesses involving massless
gluons.}.
\begin{table}[htb]
\begin{center}
\renewcommand{\arraystretch}{1.2}
\begin{tabular}{l|ll|ll}
\hline
$\phi$ & $u\bar{u}\phi$ & $\;\;\; d\bar{d}\phi$ & $ZZ\phi$ & $\;\;\; ZA\phi$ \\
\hline
$h$ & $\cos\alpha/\sin\beta$ & $-\sin\alpha/\cos\beta$ & 
$\sin(\beta-\alpha)$ & $\;\;\; \cos(\beta-\alpha)$ \\
$H$ & $\sin\alpha/\sin\beta$ & $\;\;\; \cos\alpha/\cos\beta$ & 
$\cos(\beta-\alpha)$ & $-\sin(\beta-\alpha)$ \\
$A$ & $1/\tan\beta$ & $\;\;\; \tan\beta$ & 
--- & \hspace*{7.5mm} --- \\
\hline
\end{tabular}
\caption{\it Higgs Yukawa couplings in the MSSM in units of the SM 
couplings $M_f/v$, and gauge couplings in units of the $ZZH$ 
coupling $M_Z/v$ in the SM.}
\label{tab:coeffs}
\end{center}
\renewcommand{\arraystretch}{1}
\end{table}
The parameter $\tan\beta=v_2/v_1$ is the ratio of the vacuum expectation values
of the two Higgs fields generating the masses of up- and down-type
particles, and $\alpha$ is the mixing angle in the CP-even sector. In
the decoupling limit, in which the light Higgs mass reaches the
maximum value for a given parameter $\tan\beta$, the $h$ Yukawa couplings
approach the SM values.  In
general, they are suppressed for up-type fermions and enhanced for
down-type fermions; the enhancement increases with $\tan\beta$ and can
therefore be very strong.

\begin{figure}[p]
\begin{center}
\setlength{\unitlength}{1pt}
\noindent
\begin{picture}(140,100)(15,0)
\ArrowLine(15,25)(40,50)
\ArrowLine(40,50)(15,75)
\Photon(40,50)(80,50){3}{4}
\SetWidth{1.1}
\ArrowLine(80,50)(105,75)
\ArrowLine(105,75)(130,50)
\ArrowLine(130, 0)(80,50)
\SetWidth{1}
\DashLine(105,75)(130,100){5}
\Vertex(105,75){2.5}
\put(135,45){$Q$}
\put( 98,55){$Q$}
\put(135,90){$h,H$}
\put(135, 0){$\bar{Q}$}
\put(45,35){$\gamma,Z$}
%\put(0,18){$e^-$}
%\put(0,78){$e^+$}
\end{picture}
\begin{picture}(140,100)(5,0)
\ArrowLine(15,25)(40,50)
\ArrowLine(40,50)(15,75)
\Photon(40,50)(90,50){3}{5}
\SetWidth{1.1}
\ArrowLine(90,50)(120,80)
\ArrowLine(120,20)(90,50)
\SetWidth{1}
\DashLine(65,50)(95,80){5}
\Vertex(65,50){2.5}
\put(125,80){$Q$}
\put( 90,85){$h,H$}
\put(125,15){$\bar{Q}$}
\put(45,35){$Z$}
\put(75,35){$Z$}
%\put(0,18){$e^-$}
%\put(0,78){$e^+$}
\end{picture}
\begin{picture}(150,100)(5,0)
\ArrowLine(15,25)(40,50)
\ArrowLine(40,50)(15,75)
\Photon(40,50)(80,50){3}{4}
\SetWidth{1.1}
\ArrowLine(105,75)(130,50)
\ArrowLine(130,100)(105,75)
\SetWidth{1}
\DashLine(80,50)(105,75){5}
\DashLine(130, 0)(80,50){5}
\Vertex(105,75){2.5}
\Vertex(80,50){2.5}
\put(135,45){$Q$}
\put( 98,55){$A$}
\put(135,90){$\bar{Q}$}
\put(135, 0){$h,H$}
\put(55,35){$Z$}
%\put(0,18){$e^-$}
%\put(0,78){$e^+$}
\end{picture}
\\[2em]
\begin{picture}(140,100)(15,0)
\ArrowLine(15,25)(40,50)
\ArrowLine(40,50)(15,75)
\Photon(40,50)(80,50){3}{4}
\SetWidth{1.1}
\ArrowLine(80,50)(105,75)
\ArrowLine(105,75)(130,50)
\ArrowLine(130, 0)(80,50)
\SetWidth{1}
\DashLine(105,75)(130,100){5}
\Vertex(105,75){2.5}
\put(135,45){$Q$}
\put( 98,55){$Q$}
\put(135,90){$A$}
\put(135, 0){$\bar{Q}$}
\put(45,35){$\gamma,Z$}
%\put(0,18){$e^-$}
%\put(0,78){$e^+$}
\end{picture}
\begin{picture}(150,100)(5,0)
\ArrowLine(15,25)(40,50)
\ArrowLine(40,50)(15,75)
\Photon(40,50)(80,50){3}{4}
\SetWidth{1.1}
\ArrowLine(105,75)(130,50)
\ArrowLine(130,100)(105,75)
\SetWidth{1}
\DashLine(80,50)(105,75){5}
\DashLine(130, 0)(80,50){5}
\Vertex(105,75){2.5}
\Vertex(80,50){2.5}
\put(135,45){$Q$}
\put( 92,50){$h,H$}
\put(135,90){$\bar{Q}$}
\put(135, 0){$A$}
\put(55,35){$Z$}
%\put(0,18){$e^-$}
%\put(0,78){$e^+$}
\end{picture}
\end{center}
\caption[ ]{\it 
Subchannels for the radiation of scalar and pseudoscalar MSSM Higgs
bosons in $e^+e^-$ collisions; $Q=t,b$.}
\label{fig:diags}
\end{figure}
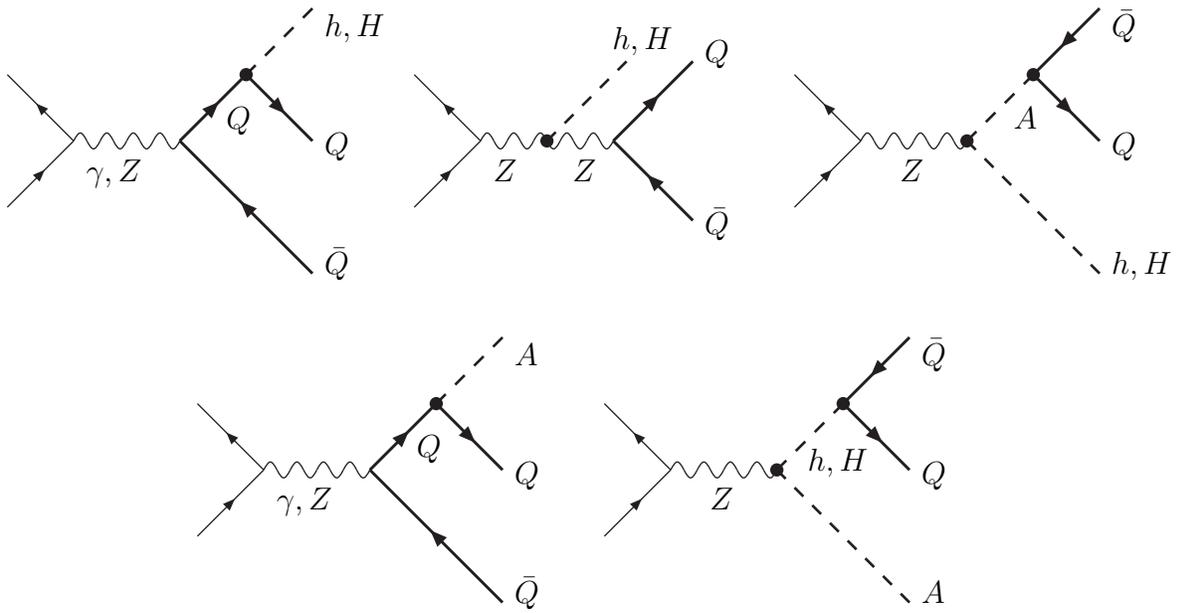

Higgs radiation off top 
or off bottom quarks in $e^+e^-$ collisions,
\begin{equation}
e^+e^- \to Q\bar{Q} h/H/A \qquad [Q=t,b],
\label{eq:processes}
\end{equation}
lends itself as a suitable process for measuring the Yukawa couplings
in supersymmetric theories \cite{dj92}, particularly for the light
Higgs boson $h$ and for moderately heavy Higgs bosons $H$ and $A$.  In
this letter we present the cross sections for these processes
including the next-to-leading order (NLO) QCD corrections. This
problem has been solved before in the SM, where the QCD corrections
significantly increase the cross section at moderate collider energies
\cite{di98,da97}. Moreover, it has been shown in detailed simulations
that the SM 
top-quark Yukawa coupling can be measured with an accuracy close to 5\%
\cite{7a} (see also Ref.~\cite{ba99}) in high-luminosity operations of $e^+e^-$
linear colliders. For the MSSM, first steps towards an NLO analysis
have been made in Ref.~\cite{da99}, adopting either a global $K$
factor from SM analyses or applying the QCD corrections to
the cross sections for double resonance
production in the narrow-width approximation. In the present analysis
we have calculated the complete set of ${\cal
O}(\alpha_\mathrm{s})$ QCD corrections to all subchannels in
(\ref{eq:processes}) systematically. The large number of interfering 
subchannels in supersymmetric theories renders this program 
more complex than
the corresponding calculation in the SM, cf.~Fig.~\ref{fig:diags}, in
particular since the relative weight of the subchannels varies over
the supersymmetric parameter space and over the phase space for
different mass ratios.
\\[2em]
{\bf 2.} Examining the set of subchannels in Fig.~\ref{fig:diags}, the QCD
corrections can be categorized into five classes. Virtual corrections
of the internal quark lines, of the $\gamma/Z$--quark vertices, of the
scalar/pseudoscalar-Higgs--quark vertices,
and box diagrams interfere with the Born amplitude. Gluon radiation
off internal and external quark lines adds incoherently to the cross
sections. In order to disentangle the different subchannels with
couplings varying over the supersymmetric parameter space, the cross
section for scalar Higgs radiation must be split into nine parts:
Higgs radiation off the quarks ($\sigma_1,\sigma_2$),
Higgs-strahlung off the $Z$-boson line ($\sigma_3,\sigma_4$), the
related interference terms ($\sigma_5,\sigma_6$), real/virtual $A$ decay
($\sigma_7$), and the interference with the first two Higgs-radiation
and Higgs-strahlung channels ($\sigma_8,\sigma_9$):
\begin{small}
\begin{eqnarray}
\lefteqn{\sigma[e^+e^- \to Q\bar{Q} H_i (g) ] = } & 
\nonumber \\[.5em]
 & & \hspace*{-0.5cm} N_{C} \frac{\sigma_0}{4\pi} \left\{ 
\frac{(\hat v_e^2 + \hat a_e^2)}{(1-M_Z^2/s)^2}
 \frac{g^2_{QQH_i}}{4\pi} 
\left( \hat v_Q^2\sigma_1 +\hat a_Q^2\sigma_2 \right)
+\left( Q_e^2Q_Q^2 + \frac{2Q_eQ_Q\hat v_e\hat v_Q}{1-M_Z^2/s}
\right) \frac{g^2_{QQH_i}}{4\pi} \sigma_1 \right. 
\nonumber \\
& & \hspace*{1cm} {} + \frac{(\hat v_e^2 + \hat a_e^2)}{(1-M_Z^2/s)^2}
\frac{g^2_{ZZH_i}}{4\pi} \left( \hat v_Q^2 \sigma_3 + \hat a_Q^2 \sigma_4
\right) 
\nonumber \\
& & \hspace*{1cm} {}+ \frac{(\hat v_e^2 + \hat a_e^2)}{(1-M_Z^2/s)^2}
\frac{g_{QQH_i}\,g_{ZZH_i}}{4\pi} 
\left( \hat v_Q^2 \sigma_5 + \hat a_Q^2 \sigma_6
\right)  + \frac{Q_eQ_Q\hat v_e\hat v_Q}{1-M_Z^2/s}
\frac{g_{QQH_i}\,g_{ZZH_i}}{4\pi} \sigma_5 
\nonumber \\
& & \hspace*{1cm} {} 
+ \frac{(\hat v_e^2 + \hat a_e^2)}{(1-M_Z^2/s)^2}
\frac{1}{4s_{2W}^2}\frac{g^2_{QQA}\,g^2_{ZAH_i}}{4\pi} \sigma_7
\nonumber \\
& & \left. \hspace*{1cm} {} + \frac{(\hat v_e^2 + \hat
a_e^2)}{(1-M_Z^2/s)^2}\frac{\hat a_Q}{2s_{2W}} \left(
\frac{g_{QQA}\,g_{ZAH_i}\,g_{ZZH_i}}{4\pi}\sigma_8
+\frac{g_{QQA}\,g_{ZAH_i}\,g_{QQH_i}}{4\pi}\sigma_9
\right)\right\},
\label{eq:eeqqh}
\end{eqnarray}
\end{small} \\[0cm]
\noindent
where $H_i = h,H$.
Correspondingly, the cross section for pseudoscalar Higgs radiation may
be written as
\begin{small}
\begin{eqnarray}
\lefteqn{\sigma[e^+e^- \to Q\bar{Q} A (g) ] = } & 
\nonumber \\[.5em]
 & & \hspace*{-0.7cm} N_{C} \frac{\sigma_0}{4\pi} \left\{ 
\frac{(\hat v_e^2 + \hat a_e^2)}{(1-M_Z^2/s)^2}
 \frac{g^2_{QQA}}{4\pi} 
\left(\hat v_Q^2 \sigma'_1 + \hat a_Q^2 \sigma'_2 \right)
+\left( Q_e^2Q_Q^2 + \frac{2Q_eQ_Q\hat v_e\hat v_Q}{1-M_Z^2/s}
\right) \frac{g^2_{QQA}}{4\pi} \sigma'_1 \right.
\nonumber \\
& & \hspace*{0.5cm} {}
+ \frac{(\hat v_e^2 + \hat a_e^2)/(4s_{2W}^2)}{(1-M_Z^2/s)^2}
\left(\frac{g^2_{QQh}\,g^2_{ZAh}}{4\pi} \sigma'_3
    + \frac{g^2_{QQH}\,g^2_{ZAH}}{4\pi} \sigma'_4
    + \frac{g_{QQh}\,g_{ZAh}\,g_{QQH}\,g_{ZAH}}{4\pi} \sigma'_5
\right)
\nonumber \\
& & \hspace*{0.5cm} {} \left.
+ \frac{(\hat v_e^2 + \hat a_e^2)}{(1-M_Z^2/s)^2}\frac{\hat a_Q}{2s_{2W}}
\left(
\frac{g_{QQA}\,g_{QQh}\,g_{ZAh}}{4\pi}\sigma'_6
+\frac{g_{QQA}\,g_{QQH}\,g_{ZAH}}{4\pi}\sigma'_7
\right)\right\},
\label{eq:eeqqa}
\end{eqnarray}
\end{small} \\[0cm]
\noindent
split into $A$ radiation off the quarks ($\sigma'_1,\sigma'_2$),
virtual Higgs decays ($\sigma'_3,\sigma'_4$) and their interference
term ($\sigma'_5$), as well as their interference terms with the
radiation diagrams ($\sigma'_6,\sigma'_7$). The Yukawa and gauge
couplings are set off explicitly in (\ref{eq:eeqqh}) and (\ref{eq:eeqqa}),
with the normalized couplings $g_{QQH_i}$ etc.\ given in 
Table~\ref{tab:coeffs}\footnote{The couplings $g_{QQ\phi}$
($g_{ZZ\phi},g_{ZA\phi}$)
include the normalization coefficients $M_Q/v$ ($M_Z/v$), respectively.}.
The electric charges are defined as $Q_e = -1$, $Q_t
= + 2/3$ and $Q_b = - 1/3$, and $\hat{v}_{e,Q}$ and $\hat{a}_{e,Q}$ are
the vector and axial-vector charges of the electron and the quark $Q$,
normalized as $\hat{v} = (I_{3L} - 2Q s^2_W) / s_{2W}$ and
$\hat{a} = I_{3L} / s_{2W}$, with $I_{3L} = \pm 1/2$ being the
weak isospin of the left-handed fermions [as usual, $s^2_W = 1 -c^2_W
= \sin^2 \theta_W = 0.23$ and $s_{2W}=\sin 2\theta_W$];
$\sigma_0=4\pi\alpha^2/3s$ denotes the standard
electromagnetic $\mu$-pair cross section,
and $N_{C} = 3$ is the color factor of the quarks.
The value of the electromagnetic coupling is taken to be $\alpha =
1/128$. The mass of the $Z$~boson is set to $M_Z=91.187\GeV$, and the
pole masses of the top and bottom quarks are set to $M_t=174\GeV$
\cite{ca98}
and\footnote{This value for the perturbative pole mass of the
bottom quark corresponds in NLO to an $\overline{\rm MS}$ mass
$\overline{m}_b(\overline{m}_b)=4.28\GeV$.}
$M_b=4.62\GeV$ \cite{bottom}, respectively. The masses of the MSSM Higgs
bosons and their couplings are related to
$\tan\beta$ and the pseudoscalar Higgs boson mass $M_A$. In
the relation we use, higher-order corrections up to two loops
in the effective-potential approach are included \cite{ca95}.
The SUSY parameters are chosen as $\mu=A_t=A_b=0$ and $M_{\tilde{Q}} =
1$ TeV; this simple choice is sufficient to illustrate the main results.

The production of $b\bar{b}h/H/A$ final states
in (\ref{eq:eeqqh}) and (\ref{eq:eeqqa})
can be mediated by resonance channels $e^+e^-\to Ah/H$ and $Zh/H$. We describe
the resonance structures as Breit--Wigner forms by
substituting $M^2\to M^2-iM\Gamma$ in all boson propagators.
The decay widths of the Higgs bosons are
calculated including higher-order corrections, as described in 
Refs.~\cite{12a,dj95}, while the $Z$ width is set to $\Gamma_Z=2.49\GeV$.
For $t\bar{t}h/H/A$ production the widths can be
neglected, since the Higgs masses are taken below the 
$t\bar t$ threshold.

The coefficients $\sigma_i$, $\sigma'_i$ in the cross sections
(\ref{eq:eeqqh}) and (\ref{eq:eeqqa}) can be decomposed into Born
contributions $\sigma^0_i$, $\sigma^{\prime 0}_i$ and QCD corrections
$\delta_i$, $\delta'_i$:
\begin{equation}
\sigma_i = \sigma^0_i 
\left(1+\frac{\alpha_{\mathrm{s}}}{\pi}\delta_i\right),
\qquad
\sigma'_i = \sigma^{\prime 0}_i 
\left(1+\frac{\alpha_{\mathrm{s}}}{\pi}\delta'_i\right).
\end{equation}
The renormalization scale of the QCD coupling $\alpha_{\mathrm{s}}$,
which is evaluated in NLO with five active flavors normalized to
$\alpha_{\mathrm{s}}(M_Z^2) = 0.119$ \cite{ca98},
is chosen at $\mu_R^2=s$, where $s=E_{\mathrm{CM}}^2$
is the center-of-mass (CM) energy squared. By definition, the Yukawa
couplings in the Born contributions are evaluated at the scale set by the
Higgs momentum flow to leading logarithmic order.

The QCD radiative corrections have been calculated in the standard
way. The Feynman diagrams and the amplitudes for the virtual
corrections have been generated with {\sl Feyn\-Arts} \cite{ku91}.
They have been evaluated by applying the standard techniques for
one-loop calculations, as described in 
Refs.~\cite{th79,de93}.
Ultraviolet divergences are consistently regularized in
$D=4-2\epsilon$ dimensions, with $\gamma_5$ treated naively since no
anomalies are involved.  The renormalization of the 
$Q\bar{Q}\phi$ vertices is
connected to the renormalization of the quark masses, which, in the case
of the top quark, is defined
on shell (see e.g.\ 
Refs.~\cite{de93,bl80} for details).  Large logarithms
are mapped into the running mass $\overline{m}_b(Q^2_{\mathrm{Higgs}})$ for the
Yukawa couplings of the $b$~quark, with $Q^2_{\mathrm{Higgs}}$ denoting
the squared momentum flow through the corresponding Higgs-boson line.
The algebraic part of the virtual
corrections has also been checked by using {\sl Feyn\-Calc}
\cite{me91}. The infrared divergences encountered in the virtual
corrections and in the cross section for 
real gluon emission, are
treated in two different ways. Both calculations follow subtraction
procedures, one using dimensional regularization and one introducing
an infinitesimal gluon mass \cite{di99}. 
The results
obtained by the two different procedures are in mutual numerical
agreement after adding the contributions from virtual gluon exchange
and real gluon emission. A second, completely independent calculation
of the QCD corrections to the total cross section was based on the
evaluation of all relevant cut diagrams of the photon and $Z$-boson
self-energies in two-loop order, generalizing the method applied to
$t\bar{t}(g)$ intermediate states in Ref.~\cite{dg94}. 
The results of the two approaches are in numerical agreement.
The final expressions for the integrated cross sections $\sigma^0_i$,
$\sigma^{\prime 0}_i$ and for the QCD corrections $\delta_i$, $\delta'_i$ 
are, however, too lengthy to be recorded in this letter\footnote{NLO
Fortran programs calculating the integrated cross sections and 
differential distributions for all the processes, are available on the web
\cite{WWW}.}.
\\[2em]
{\bf 3.} The general characteristics of the reactions (\ref{eq:processes})
will be discussed for two examples in detail, $\tan\beta=3$ and 30.
The numerical analysis has been performed
for the case of no mixing in the stop and sbottom sectors of the MSSM.
Since the top Yukawa couplings are suppressed for large values of
$\tan\beta$, sizeable cross sections for $t\bar t\phi$ production are
expected only for small and moderate values of $\tan\beta$ and in the
decoupling regime at large values of $\tan\beta$,
where the $h$
couplings approach the SM values. The opposite is realized for
$b\bar b\phi$ production outside the resonance regions due to the strong
enhancement of the $b$ Yukawa couplings for large values of $\tan\beta$.

{\bf a.)}
The QCD corrections to the top final states $t\bar t\phi$ can be interpreted
easily in two kinematical areas. Whenever the invariant mass of the $t\bar
t$ pair is close to 
threshold, the gluonic Sommerfeld rescattering-correction
is positive and becomes large. In the threshold region the $K$
factor approaches the asymptotic form \cite{di98}
\begin{equation}
K^{t\bar t\phi}_{\mathrm{thr}} \to 1+\frac{32\alpha_s}{9\beta_t}
\label{eq:Coulsing}
\end{equation}
with the maximal quark velocity 
$\beta_t=\sqrt{(\sqrt{s}-M_\phi)^2-4 M_t^2}/2M_t$ 
%$\beta_t=\sqrt{(\sqrt{}s-M_\phi)^2-4 M_t^2}/2M_t$ 
in the $(t\bar t)$ rest frame.

For high energies, on the other hand, the QCD corrections are of order
$\alpha_s/\pi$. In the energy region $s\gg 4M_t^2\gg M_H^2$ but $\log
s/M_t^2 \not \!\gg {\cal O}(1)$, which is relevant for the present
analysis, the QCD corrections can qualitatively
be traced back to vertex corrections and infrared gluon radiation. Since
scalar Higgs bosons are radiated off top quarks preferentially with
small energy [$x=E_\phi/E_t \to 0$], as is evident from the leading
(universal) part of the fragmentation function
\begin{equation}
f(t\to th/H;x) = \frac{g^2_{tth/H}}{16\pi^2} \left[ 4~\frac{1-x}{x} +
x~\log \frac{s}{M_t^2} \right] \,\, ,
\label{eq:fragh}
\end{equation}
the QCD correction of the scalar Yukawa vertex, regularized by soft
gluon radiation, approaches the value \cite{da97}
\begin{equation}
\Delta^{V+IR}_{h/H} = \frac{4\alpha_s}{3\pi} \left[ -1+ \frac{2-x}{x}
\log(1-x) \right] \to -4\frac{\alpha_s}{\pi} \,\, .
\end{equation}
The scalar Yukawa vertex is
therefore reduced by four units in $\alpha_s/\pi$ which are compensated
only partly by one unit due to the increase of the $t\bar t$ production
probability, leading in total \cite{di98,da97} to
\begin{equation}
K^{t\bar t\phi}_{\mathrm{cont}} \to 1-3\frac{\alpha_s}{\pi} + \ldots
\hspace*{0.5cm} \mbox{for $\phi=h,H$} \,\, .
\end{equation}
The ellipsis accounts for hard Higgs and gluon radiation (of order
$+\alpha_s/\pi$). Thus, the QCD corrections are expected negative for
scalar Higgs particles in the high energy continuum.

By contrast, the corresponding fragmentation function for the pseudoscalar
Higgs boson \cite{beenakker}
\begin{equation}
f(t\to tA;x) = \frac{g^2_{ttA}}{16\pi^2} ~x~\log \frac{s}{M_t^2}
\label{eq:fraga}
\end{equation}
is hard so that the average of the vertex and IR gluon corrections over
the Higgs spectrum amounts to
\begin{equation}
\Delta^{V+IR}_A \to \frac{4\alpha_s}{3\pi} \left\langle \left[ 1+ \frac{2-x}{x}
\log(1-x) \right] \right\rangle \sim -\frac{3}{2} \frac{\alpha_s}{\pi} \,\, .
\end{equation}
Adding to this correction the increase of the $t\bar t$ production
probability of one unit, the $K$ factor is very close to unity
\begin{equation}
K^{t\bar tA}_{\mathrm{cont}} \to 1-\frac{1}{2}\frac{\alpha_s}{\pi} + \ldots
\end{equation}
After hard gluon bremsstrahlung is taken into account (symbolized by the
ellipsis), the overall QCD corrections for the pseudoscalar Higgs boson
are therefore expected slightly positive. [For ultra-high energies,
i.e.~$\log s/M_t^2\gg 1$, hard gluon bremsstrahlung becomes important.
Similarly to the leading terms in the fragmentation functions
eqs.~(\ref{eq:fragh}) and (\ref{eq:fraga}), the QCD corrections for
scalar and pseudoscalar Higgs bosons approach each other as a result of
chiral symmetry restoration in {\it asymptotia}; this has been verified
in a numerical calculation.]

Similar estimates can be applied to bottom final states which in general are
dominated by resonance decays. After absorbing the large logarithms
$\log(Q^2_\phi/M_b^2)$ into the Yukawa couplings, the non-leading effects
are positive:
\begin{equation}
K^{b\bar b\phi}_{\mathrm{res}} \approx 1+\left\{{\textstyle \frac{17}{3}},1
\right\} \frac{\alpha_s}{\pi} 
\hspace*{0.5cm}
\mbox{for $\left\{ \mbox{Higgs}, Z\right\} \to b\bar b$ } \,\, .
\end{equation}
Also close to the thresholds and in the high-energy limit the QCD
corrections remain positive after mapping the large (negative) corrections
into the running Yukawa couplings. Since different channels are activated at the
same time, only a qualitative estimate can be given in the continuum
regime,
\begin{equation}
K^{b\bar b\phi}_{\mathrm{cont}} = 1+c\frac{\alpha_s}{\pi} 
\hspace*{0.5cm}
\mbox{with $c = {\cal O}(1)$,}
\end{equation}
while details must be left to the numerical analysis.

\begin{figure}[hbtp]
\vspace*{0.0cm}

\hspace*{-1.7cm}
\begin{turn}{-90}%
\epsfxsize=7cm \epsfbox{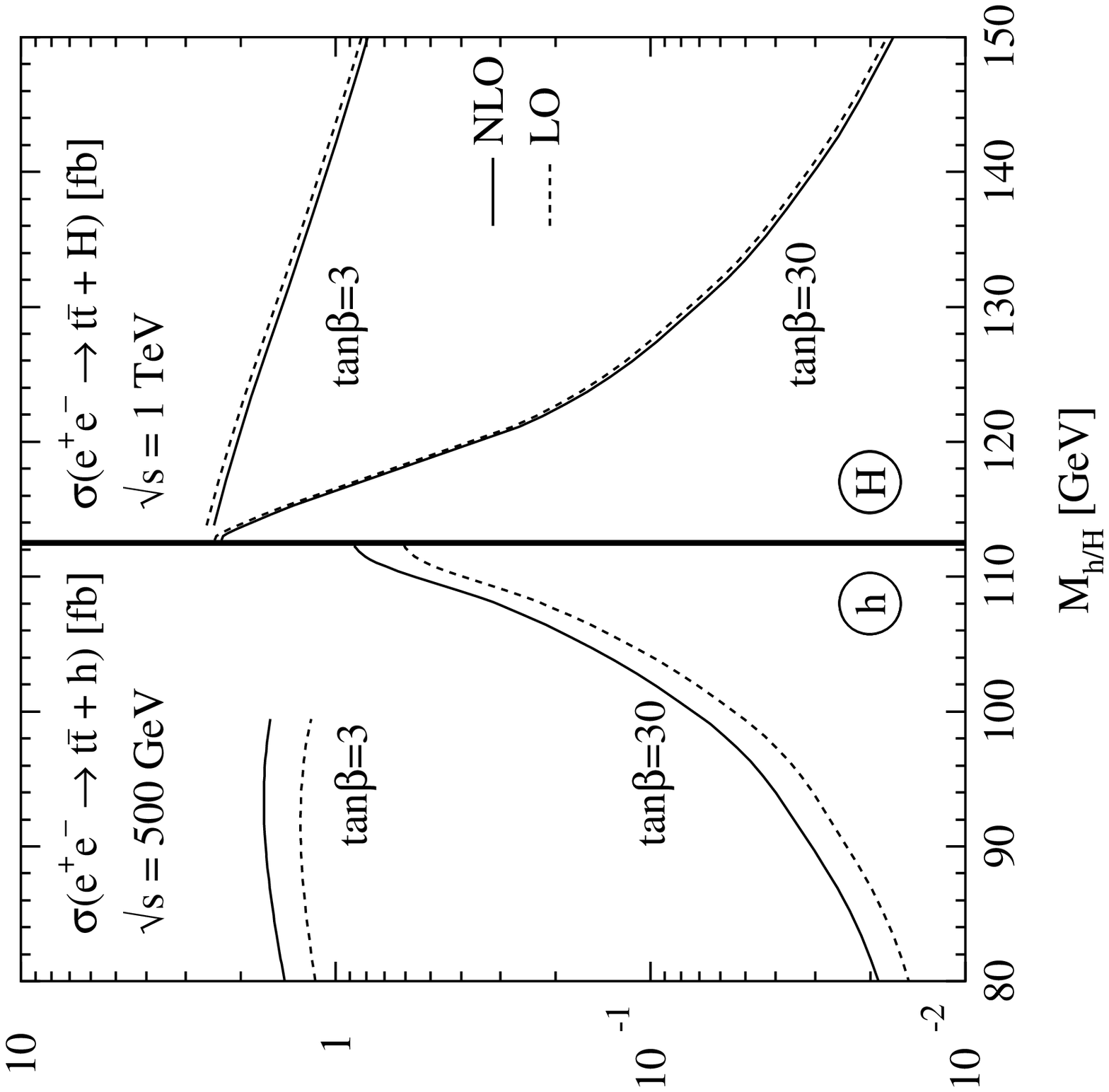}
\end{turn}
\hspace*{-1.8cm}
\begin{turn}{-90}%
\epsfxsize=7cm \epsfbox{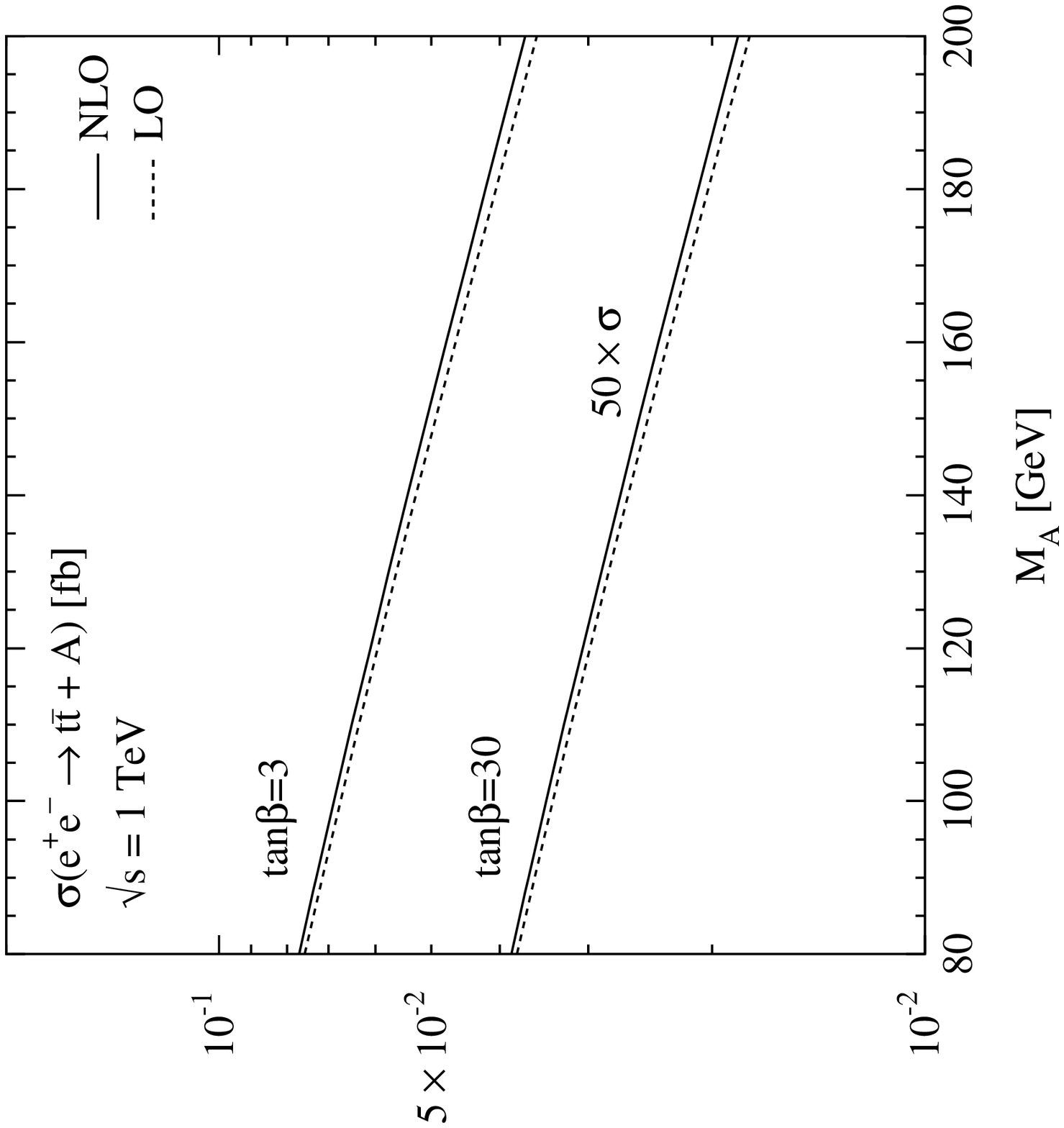}
\end{turn}
\vspace*{0.5cm}

\hspace*{-1.7cm}
\begin{turn}{-90}%
\epsfxsize=7cm \epsfbox{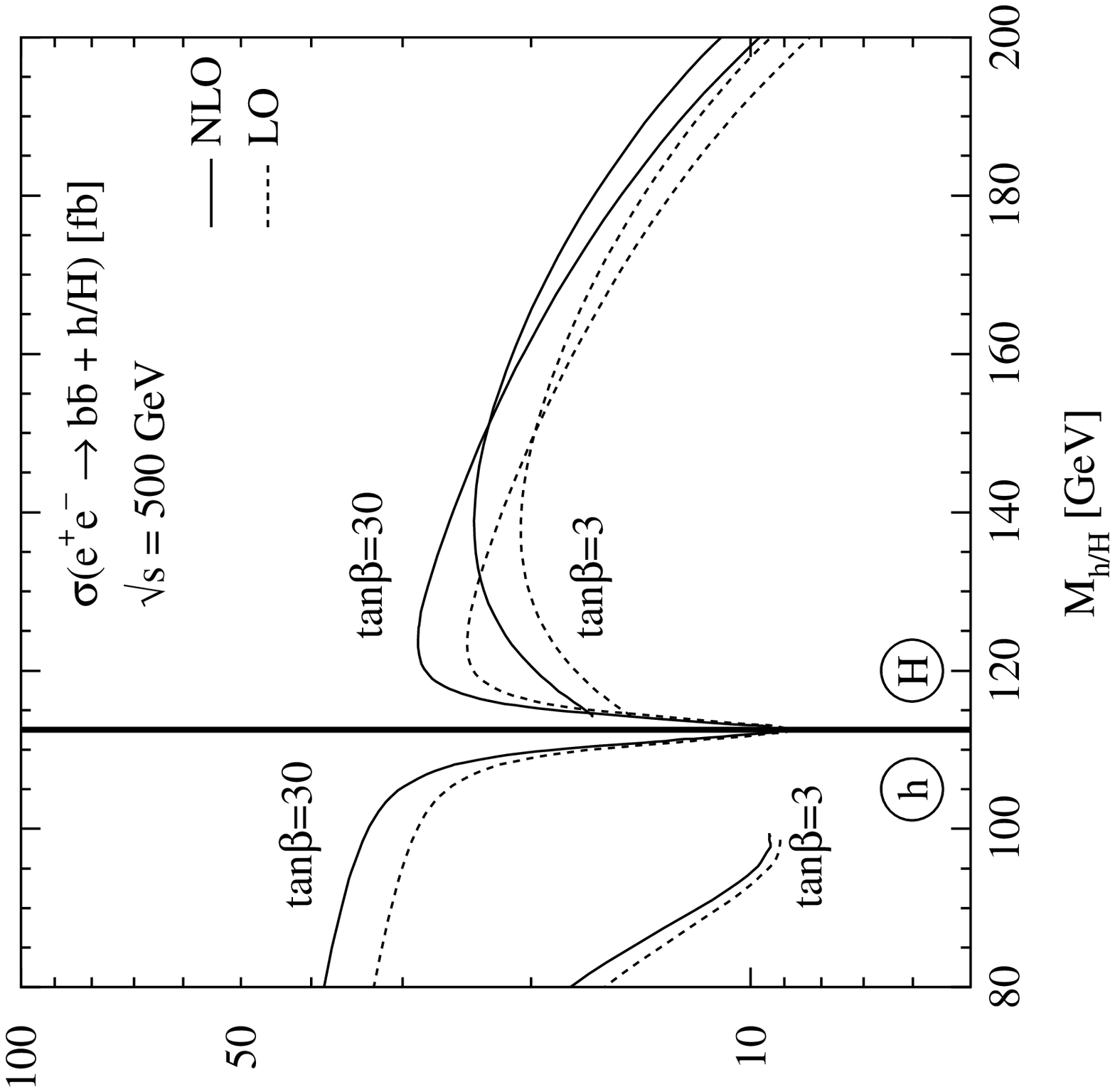}
\end{turn}
\hspace*{-1.8cm}
\begin{turn}{-90}%
\epsfxsize=7cm \epsfbox{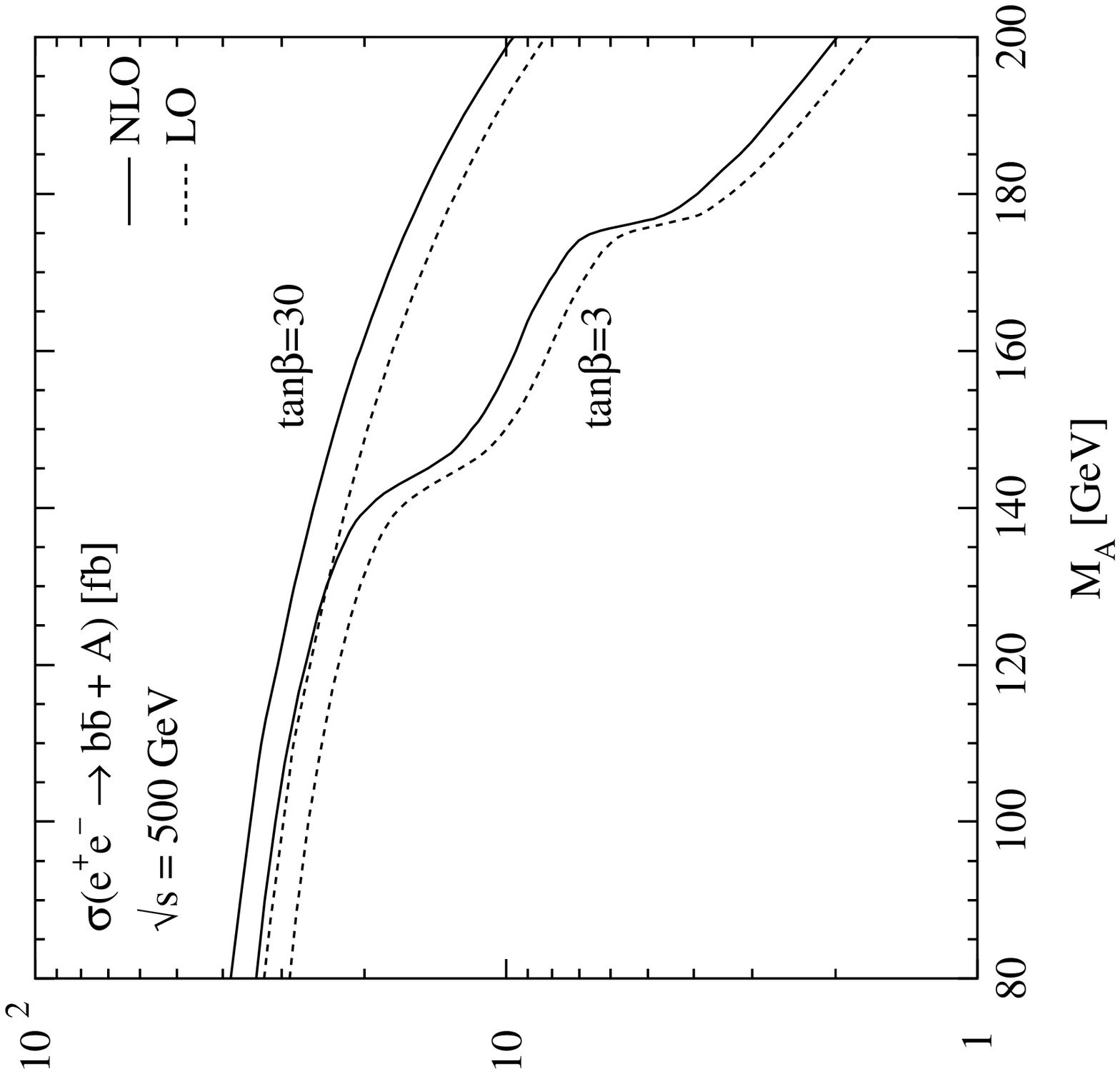}
\end{turn}
\vspace*{0.5cm}

\caption[ ]{\it 
Production cross sections for the MSSM Higgs bosons $h,H$ and $A$ in
association with heavy $t,b$ quark pairs;
Born approximation: dashed, QCD corrections included: full curves. The rapid
drops in the $b\bar bA$ cross section at $\tan\beta=3$ are due to the
kinematical opening of the resonant $H\to WW, hh$ decays, which reduce
the branching ratio of the resonant $H\to b\bar b$ decay.
\label{fig:cs}}
\end{figure}

{\bf b.)}
The results are exemplified in Fig.~\ref{fig:cs} for $\tan\beta=3$ and
30 and for the collider energy $E_{\mathrm{CM}}=500\GeV$.
If required by the size of the cross section, which should not fall
below $\sim 10^{-2}$ fb in order to be accessible experimentally, we switched
to $E_{\mathrm{CM}}=1\TeV$.
The Born
terms are shown by the dotted curves, while the final results for the cross
sections, including QCD corrections, are given by the full curves.

For $E_{\mathrm{CM}}=500\GeV$ 
the QCD corrections to Higgs-boson production in
association with $t\bar t$ pairs increase the scalar Higgs-production
cross sections significantly,
as can be inferred from
Fig.~\ref{fig:cs}. Close to threshold the numerical results
clearly exhibit the strong increase of the cross sections due to the
Coulomb singularity (\ref{eq:Coulsing}).
Moreover, for $\tan\beta=30$ the cross sections
are strongly suppressed except for the regions 
where the light (heavy)
scalar Higgs mass is close to its upper (lower) bound. For
$\tan\beta=3$ the cross section amounts to about 1 fb, which
leads to a significant number of events at the TESLA collider,
being designed to reach integrated luminosities of about $\int
{\cal L} \sim 1$ ab$^{-1}$ in three years of operation.

For CM energies of 1 TeV the QCD corrections to scalar Higgs
production in association with $t\bar t$ pairs are of moderate size.
They decrease the cross sections by about 3--5\%. This is in
accordance with the asymptotic form of the $K$ factor \cite{di98}.
For pseudoscalar Higgs production, the size of the QCD corrections is
slightly positive at 1 TeV, in agreement with the qualitative
discussion above.

The cross sections for
Higgs-boson production associated with $b\bar b$
pairs are significantly larger due to the resonance contributions from on-shell
$Z$ and 
Higgs-boson decays into $b\bar b$ pairs. The QCD corrections
increase these cross sections by about 
5--25\%. The drop in the $b\bar bA$ production cross
section for $\tan\beta=3$ at $M_A\sim 140\GeV$ and $175\GeV$ can be attributed
to the crossing of the thresholds for resonant $H\to WW$ and $H\to hh$
decays, respectively, in $HA$ final states.

\begin{figure}[hbtp]
\vspace*{0.0cm}

\hspace*{-1.7cm}
\begin{turn}{-90}%
\epsfxsize=7cm \epsfbox{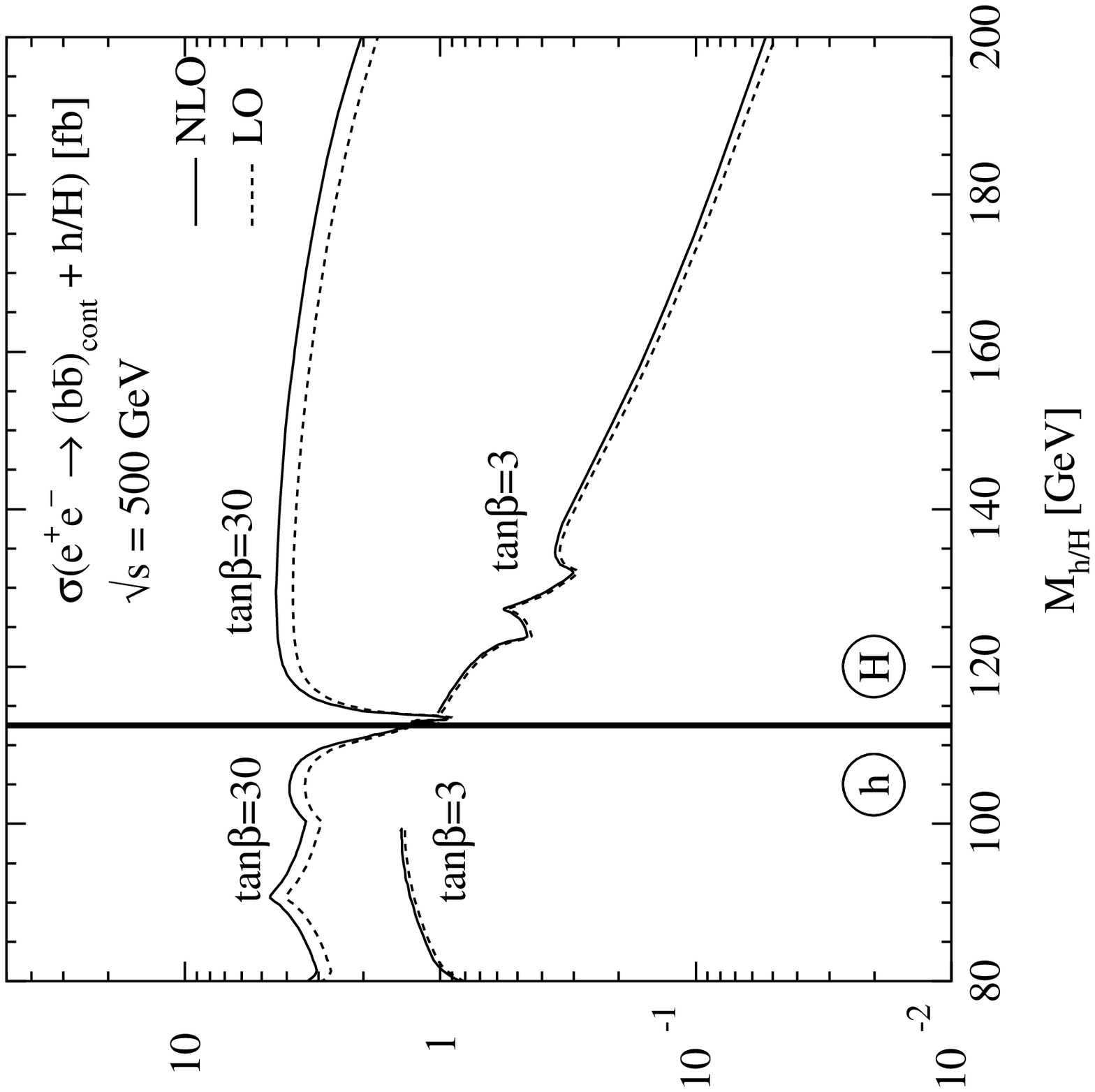}
\end{turn}
\hspace*{-1.8cm}
\begin{turn}{-90}%
\epsfxsize=7cm \epsfbox{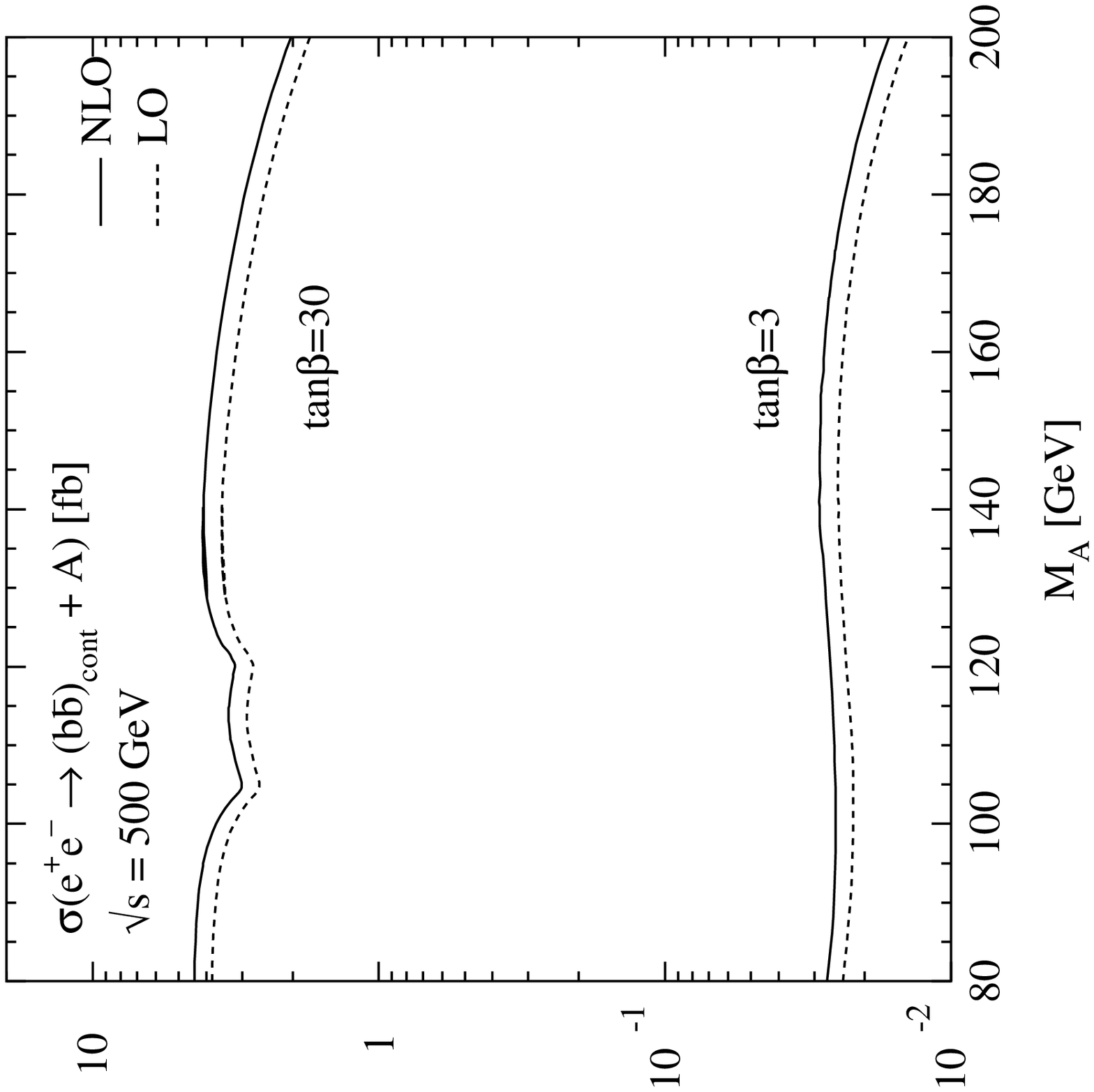}
\end{turn}
\vspace*{0.5cm}

\caption[ ]{\it Continuum production of MSSM Higgs bosons $h,H,A$ in
association with a $b\bar b$ pair after subtraction of resonance decays
to $b\bar b$ pairs in the Breit-Wigner bands $M_R\pm 5$ GeV.
\label{fg:cont}}
\end{figure}

Without cuts, 
the intermediate resonance decays $Z,h,H,A\to b\bar b$ dominate all
$b\bar b\phi$ production processes, whenever they are kinematically
allowed, and it will be difficult to extract the bottom Yukawa
couplings experimentally in regions where resonant Higgs decays
to $b\bar b$ pairs are dominant. This is the case at large values of
$\tan\beta$ for all neutral Higgs particles and at small values of
$\tan\beta$ for Higgs masses below the $WW$ threshold. In these cases
the branching ratios, which determine the size of the $b\bar b\phi$ 
cross sections, will be 
nearly independent of the bottom Yukawa couplings.
Resonance decays 
$R\to b\bar b$ pairs in the $b\bar b h/H/A$ final states can however
be eliminated by cutting out the resonance energy of the 
final-state Higgs boson, 
$E_{\phi,\mathrm{res}} = (s+M^2_\phi-M_R^2)/2\sqrt{s}$.
After subtracting these resonance parts, the non-resonant
contributions are suppressed by about one to three orders of magnitude.
The resonances have been removed from the cross sections in the examples of
Fig.~\ref{fg:cont} by subtracting the two-particle
cross sections in the Breit--Wigner bands $M_R\pm \Delta$ of the energy
$E_{\phi,\mathrm{res}}$ with
the resolution $\Delta = 5\GeV$.
This theoretical definition is used for the sake of simplicity; wider cuts
may be required in experimental analyses.
Peaks and dips in the cross sections are the result of overlapping
Breit-Wigner bands. For the scalar Higgs
particles they arise from overlapping $Z$ and $A$ boson bands; in
pseudoscalar Higgs production the light and heavy scalar
resonance bands overlap for 100 GeV $\lsim M_A \lsim 120$ GeV for
$\tan\beta=30$. The dips occur whenever the two resonance bands
touch each other, while the peaks between the dips occur when the
resonance masses coincide exactly. As shown in Fig.~\ref{fg:cont}, the 
QCD-corrected cross sections are still close to $1\fb$ or slightly below,
except for heavy masses at small values of $\tan\beta$.
Thus, ensembles of order $10^3$ events can be collected at a high-luminosity
collider.
\\[2em]
{\bf 4.} Measurements of Yukawa couplings in supersymmetric Higgs radiation
off heavy quarks at $e^+e^-$ linear colliders are difficult. This is a
result of the large number of subchannels contributing to the $Q\bar Qh/H/A$
final states in supersymmetric theories in general, and the contamination by
two-Higgs final states in particular. Nevertheless, the continuum cross
sections appear large enough to approach this experimental problem as
shown in the present analysis. Even though experimental simulations are beyond the
scope of this note, it may be concluded from earlier SM analyses that the
method will work at least in part of the MSSM parameter space, thus providing us
with the absolute size of the quark--Higgs Yukawa couplings in the minimal
supersymmetric theory.

\mbox{}
\\[2em]
{\small {\bf Acknowledgement}\\
Y.L.\ thanks Prof.~A.~Wagner for the warm hospitality extended to him 
during his visit at DESY, based on the DESY--Tsinghua cooperation
contract.}

\clearpage

\end{document}